\PassOptionsToPackage{unicode}{hyperref}
\PassOptionsToPackage{hyphens}{url}
\documentclass[
  11pt,
]{article}
\usepackage{amsmath,amssymb}
\usepackage{iftex}
\ifPDFTeX
  \usepackage[T1]{fontenc}
  \usepackage[utf8]{inputenc}
  \usepackage{textcomp} 
\else 
  \usepackage{unicode-math} 
  \defaultfontfeatures{Scale=MatchLowercase}
  \defaultfontfeatures[\rmfamily]{Ligatures=TeX,Scale=1}
\fi
\usepackage{lmodern}
\ifPDFTeX\else
\fi
\IfFileExists{upquote.sty}{\usepackage{upquote}}{}
\IfFileExists{microtype.sty}{
  \usepackage[]{microtype}
  \UseMicrotypeSet[protrusion]{basicmath} 
}{}
\makeatletter
\@ifundefined{KOMAClassName}{
  \IfFileExists{parskip.sty}{%
    \usepackage{parskip}
  }{
    \setlength{\parindent}{0pt}
    \setlength{\parskip}{6pt plus 2pt minus 1pt}}
}{
  \KOMAoptions{parskip=half}}
\makeatother
\usepackage{xcolor}
\usepackage[margin=2.8cm]{geometry}
\usepackage{longtable,booktabs,array}
\usepackage{calc} 
\usepackage{etoolbox}
\makeatletter
\patchcmd\longtable{\par}{\if@noskipsec\mbox{}\fi\par}{}{}
\makeatother
\IfFileExists{footnotehyper.sty}{\usepackage{footnotehyper}}{\usepackage{footnote}}
\makesavenoteenv{longtable}
\providecommand{\tightlist}{%
  \setlength{\itemsep}{0pt}\setlength{\parskip}{0pt}}
\ifLuaTeX
  \usepackage{selnolig}  
\fi
\IfFileExists{bookmark.sty}{\usepackage{bookmark}}{\usepackage{hyperref}}
\IfFileExists{xurl.sty}{\usepackage{xurl}}{} 
\hypersetup{
  pdftitle={Software Delegation Contracts: Measuring Reviewability in AI Coding-Agent Work},
  hidelinks,
  pdfcreator={LaTeX via pandoc}}

\title{Software Delegation Contracts: Measuring Reviewability in AI
Coding-Agent Work}
\author{Vincent Schmalbach\\
Independent Researcher\\
vschmalbach@vschmalbach.com\\
https://www.vincentschmalbach.com/}
\date{June 2026}

\begin{document}
\maketitle
\begin{abstract}
AI coding agents increasingly work in a delegated mode: they accept an
assigned software task, act inside a repository under bounded authority,
and return a work package for human review. Prior conceptual work
proposed the \emph{software delegation contract}, defined here as the
task, authority, work package, and acceptance context around such a run,
as the unit of analysis for delegated coding work, but offered no
empirical evidence. This paper reports a controlled pilot study of that
claim. We built an instrumented task environment (a small
dependency-free TypeScript API with seeded defects and documentation
gaps), authored ten tasks across five families, and executed 64
coding-agent runs across two model tiers under three conditions: a
realistic issue-style prompt, an explicit delegation contract, and a
contract with a required evidence bundle. Every run was scored
mechanically (hidden acceptance tests, mutation checks, scope analysis)
and reviewed by three independent, condition-blinded model-based
reviewers using a fixed rubric (192 reviews). Explicit contracts did not
change objective task outcomes, which saturated: all 64 runs passed
hidden acceptance checks with zero scope violations. They did change
reviewability. Evidence sufficiency improved in 22 of 30 paired
comparisons and worsened in none (+0.83 on a 5-point scale, p
\textless{} 0.0001, Cliff's delta = 0.66); reviewer ambiguity decreased
(p = 0.035); structured work-package elements such as changed-file lists
(7\% to 93\%) and known-limitations sections (0\% to 80\%) appeared
almost exclusively under contracts; and reviewer-checklist and
residual-risk sections appeared only when the contract demanded them.
Contracts cost +13\% agent tokens and +38\% wall-clock time. The effect
was larger for the weaker model tier. We conclude that on small tasks
with capable models, delegation contracts buy reviewability rather than
correctness, and we derive concrete defaults for harness builders. The
task environment, prompts, patches, reports, reviews, and analysis
scripts are preserved as a companion evidence artifact.
\end{abstract}

\hypertarget{introduction}{%
\section{Introduction}\label{introduction}}

AI coding agents now go beyond interactive code completion. Contemporary
systems accept assigned software tasks, inspect repositories, run tests,
modify files, and return pull requests or patches with explanatory
reports {[}1, 12, 13{]}. We use the practitioner term \textbf{software
delegate} for a coding agent used in this mode: assigned software work,
bounded authority, and a reviewable artifact returned for acceptance or
rejection.

When work is delegated, the binding question for the human is no longer
only \emph{does the patch work?} but \emph{can I review this?} A
reviewer must know what was asked, what the agent was allowed to do,
what came back, and what evidence supports it. Prior conceptual work,
including our own, argued that the right unit of analysis for this
relationship is the \textbf{delegation contract}: the task, authority,
work package, and acceptance context around a coding-agent run {[}4,
5{]}. That argument was conceptual; it produced definitions and a coding
scheme, not measurements.

This paper reports the first controlled measurements and asks one narrow
question:

\begin{quote}
When a human delegates a software task to a coding agent, does making
the delegation contract explicit improve the reviewability of the
returned work package, and at what cost?
\end{quote}

We answer it with a paired pilot study: ten seeded tasks across five
families in a purpose-built toy repository, executed by two model tiers
under (A) a realistic issue-style prompt, (B) an explicit delegation
contract, and (C) a contract plus a required evidence-bundle template,
for 64 runs in total. Outcomes are measured two ways:
\emph{mechanically}, against hidden acceptance tests, mutation checks,
and authority boundaries the agent never sees; and \emph{by review},
with three independent, condition-blinded, model-based reviewers scoring
every work package on a fixed rubric (192 reviews).

The measurements separate outcome quality from reviewability:

\begin{enumerate}
\def\labelenumi{\arabic{enumi}.}
\tightlist
\item
  \textbf{Objective outcomes saturated.} All 64 runs passed the hidden
  acceptance checks; no run violated its authority boundary. On small,
  well-specified tasks, current models do not need a contract to
  succeed.
\item
  \textbf{Reviewability changed.} Evidence sufficiency rose in 22 of 30
  paired comparisons and fell in none (+0.83 on a 5-point scale; p
  \textless{} 0.0001; Cliff's \(\delta\) = 0.66). Reviewer ambiguity
  fell (p = 0.035, judge-level p = 0.0001). Structured report elements
  appeared almost exclusively under contracts, and some elements
  (residual risks, reviewer checklists) appeared \emph{only} when the
  contract's evidence bundle demanded them.
\item
  \textbf{Contracts added run cost.} Agent tokens rose 13\%, wall-clock
  time rose 38\%, and tool invocations rose 23\% (all p \(\leq\) 0.001).
  That overhead bought evidence, not correctness.
\item
  \textbf{The weaker model benefited more}, suggesting contracts
  partially substitute for the reporting discipline stronger models
  exhibit unprompted.
\end{enumerate}

Contributions. (i) A reusable experimental harness for
delegation-contract studies, built from a seeded task repository, paired
prompts, hidden acceptance and mutation checks, mechanical scope
analysis, and a blinded review pipeline. (ii) The first paired empirical
comparison of prompt-only versus contract-based delegation to coding
agents, with 64 runs and 192 blinded reviews. (iii) Evidence that
delegation contracts function as a \emph{reviewability} mechanism rather
than a \emph{correctness} mechanism on small tasks, with quantified
cost. (iv) Design implications for agent harnesses, including which
work-package fields appear only when demanded and what they cost.

\hypertarget{background-and-definitions}{%
\section{Background and Definitions}\label{background-and-definitions}}

\textbf{Delegation contract.} Following the conceptual framework
{[}4{]}, a delegation contract is a tuple \(\langle T, A, W, C\rangle\):
the \textbf{task} T assigned to the agent (objective, scope, non-goals,
success criteria); the \textbf{authority} A granted to it (files it may
modify, commands it may run, actions that are forbidden); the
\textbf{work package} W it returns (artifact plus evidence: changed
files, commands, test results, limitations); and the \textbf{acceptance
context} C through which the work enters a delivery process (who
reviews, against what criteria). A \textbf{software delegate} is a
coding agent operating under such a contract. In most current practice
the contract is implicit, assembled from a prompt, repository
permissions, sandbox behavior, and review habits. The experimental
manipulation in this study is precisely whether the contract is made
explicit.

\textbf{Supervisory control.} Supervisory-control theory treats
automation as something that can be applied at different levels to
information acquisition, analysis, decision selection, and action
implementation {[}6{]}. A delegation contract is a policy for allocating
automation across these functions and defining the evidence the human
side of the loop needs. The present study holds the automation profile
fixed (full bounded execution, review after return) and varies only the
explicitness of the contract.

\textbf{Reviewability.} We define reviewability with a reviewer-facing
rubric: scope compliance, evidence sufficiency, reviewer confidence,
ambiguity, repair required, and an acceptance decision (Section 3.6).
Human-AI teaming research holds that the value of an AI teammate depends
on verification cost as well as raw capability {[}7{]}.

\hypertarget{method}{%
\section{Method}\label{method}}

\hypertarget{typescript-http-api-task-environment}{%
\subsection{TypeScript HTTP API task
environment}\label{typescript-http-api-task-environment}}

We built a small TypeScript HTTP API (\texttt{toy-ts-api},
\textasciitilde600 lines) with users/notes routes, an HMAC-token auth
helper, validation and slug helpers, and a 29-test suite using Node's
built-in test runner. The repository has zero external dependencies, so
every clone is immediately testable. It contains no private or product
code.

Starting from a clean base commit with all tests green, we created one
seeded branch per task:

\begin{longtable}[]{@{}
  >{\raggedright\arraybackslash}p{(\columnwidth - 6\tabcolsep) * \real{0.1304}}
  >{\raggedright\arraybackslash}p{(\columnwidth - 6\tabcolsep) * \real{0.1739}}
  >{\raggedright\arraybackslash}p{(\columnwidth - 6\tabcolsep) * \real{0.3043}}
  >{\raggedright\arraybackslash}p{(\columnwidth - 6\tabcolsep) * \real{0.3913}}@{}}
\toprule\noalign{}
\begin{minipage}[b]{\linewidth}\raggedright
Task
\end{minipage} & \begin{minipage}[b]{\linewidth}\raggedright
Family
\end{minipage} & \begin{minipage}[b]{\linewidth}\raggedright
Seeded state
\end{minipage} & \begin{minipage}[b]{\linewidth}\raggedright
Visible failures
\end{minipage} \\
\midrule\noalign{}
\endhead
\bottomrule\noalign{}
\endlastfoot
T01 & failing-test fix & token expiry compares seconds to milliseconds &
2 \\
T02 & failing-test fix & page-size clamp removed from notes route & 1 \\
T03 & validation bug & email regex silently rejects plus-addressing & 0
(latent) \\
T04 & validation bug & date-range filter accepts reversed ranges & 0
(latent) \\
T05 & missing tests & no tests for documented slug rules & 0 (gap) \\
T06 & missing tests & no test for tampered-token rejection & 0 (gap) \\
T07 & scoped refactor & pagination parsing duplicated in two routes &
0 \\
T08 & scoped refactor & misleadingly named \texttt{formatDate} function
& 0 \\
T09 & docs update & setup docs reference a renamed npm script & 0 \\
T10 & docs update & supported environment variables undocumented & 0 \\
\end{longtable}

Each task record lists the seed commit, the allowlist of files the agent
is expected to touch, frozen paths, the expected test command, and
private ground-truth notes. The seeded bugs are realistic regressions: a
unit mismatch introduced by a ``refactor'' and a ``tightened''
validation regex, both committed with plausible messages.

\hypertarget{experimental-conditions}{%
\subsection{Experimental conditions}\label{experimental-conditions}}

\textbf{A. Baseline prompt.} A realistic issue-style assignment, as a
competent colleague would write it (``Bug report: a user can't sign up
with \texttt{pat+newsletter@example.com} \ldots{} Please fix this
without breaking existing validation.''). Baselines are deliberately
\emph{good} prompts, including natural scope hints, to avoid a straw-man
comparison.

\textbf{B. Explicit delegation contract.} A structured document with the
sections: Objective, Non-goals, Authority (allowed paths, allowed
commands, forbidden actions), Expected tests, Required evidence
(summary, changed files with reasons, commands run, tests and results,
known limitations), and Acceptance criteria. Task content is
informationally equivalent to A; the manipulation is the explicit
contract structure, scope boundary, and evidence requirement.

\textbf{C. Contract plus evidence bundle.} Condition B plus a mandatory
work-package template with fixed sections (Summary, Changed files,
Commands run, Tests and results, Known limitations, Residual risks,
Reviewer checklist). Run as a four-run preview (T01 and T03 on both
model tiers), per the staged pilot design.

\hypertarget{claude-code-agent-runs}{%
\subsection{Claude Code agent runs}\label{claude-code-agent-runs}}

Autonomous coding subagents ran in the Claude Code harness on two model
tiers: \textbf{Sonnet 4.6} (stronger) and \textbf{Haiku 4.5}
(faster/weaker). Each run received a fresh single-branch clone of the
task branch and a fixed harness preamble: work only in the clone, no
commits, no package installation, no network, \texttt{npm\ test}
allowed, and the final reply was the only report delivered. The agent's
final message is the \textbf{work package}; the working tree is its
artifact. Agents were not told they were in a study.

The full grid comprised 10 tasks × \{A, B\} × \{Sonnet, Haiku\}
(replicate 1) + 10 tasks × \{A, B\} × Sonnet (replicate 2) + 4
C-previews, yielding \textbf{64 runs} and \textbf{30 A/B pairs} matched
on (task, model, replicate).

\hypertarget{automated-scoring}{%
\subsection{Automated scoring}\label{automated-scoring}}

After each run, an automated collector computed, without any model in
the loop: changed files (vs.~the seed commit) and scope violations
against the task's allowlist; forbidden-action flags (dependency
changes, \texttt{node\_modules}, commits); visible-suite results; and
\textbf{hidden acceptance checks} the agent never saw:

\begin{itemize}
\tightlist
\item
  \emph{Hidden tests} (T01--T04, T07, T08): independently written
  acceptance tests copied into the clone post-hoc, exercising the
  ground-truth behavior (e.g., reversed ranges return 400; the rename
  preserves behavior and removes the old export).
\item
  \emph{Mutation checks} (T05, T06): the collector injects a defect the
  existing suite provably does not catch (slug collision suffixing
  disabled; token signature verification removed) and requires the
  agent's new tests to fail under the mutant.
\item
  \emph{Static checks} (T09, T10): only documentation files changed and
  the documented commands/variables match the code.
\end{itemize}

\hypertarget{blinded-review}{%
\subsection{Blinded review}\label{blinded-review}}

Every work package was reviewed by \textbf{three independent model-based
reviewers} (Sonnet 4.6) in three passes. Each pass shuffled the 64 items
with a different seed, assigned fresh blind item codes, and grouped
items into batches of eight per reviewer instance. Reviewers saw the
\emph{neutral issue-form task description} (identical across
conditions), the diff, the contributor's report with revealing paths
redacted, and the visible CI result, but never the condition label, the
hidden acceptance results, or the contract text. Reviewers scored the
fixed rubric: scope compliance (1--5), evidence sufficiency (1--5),
reviewer confidence (1--5), ambiguity (1--5, lower better), repair
required (0/1), a four-level decision, and free-text notes. We aggregate
per-run scores as the median of the three reviewers. Review \emph{time}
was not measured; model reviewers have no meaningful minutes (Section
6).

\hypertarget{statistical-analysis}{%
\subsection{Statistical analysis}\label{statistical-analysis}}

Primary analysis is paired across the 30 A/B pairs: Wilcoxon signed-rank
(zeros dropped, tie-corrected normal approximation), an exact sign test,
and Cliff's \(\delta\) on the pooled distributions. A secondary
judge-level analysis pairs the three reviewers' sorted scores within
each A/B cell (90 paired scores per metric). Given the pilot scale we
treat p-values as descriptive strength-of-signal indicators, not
confirmatory thresholds. The hypotheses follow the prior framework's
testing agenda {[}4{]}: H1 ambiguity ↓, H2 evidence sufficiency ↑, H3
authority violations ↓, H4 cost ↑ with offsetting review benefit.

\hypertarget{results}{%
\section{Results}\label{results}}

\hypertarget{objective-outcomes-reach-a-ceiling}{%
\subsection{Objective outcomes reach a
ceiling}\label{objective-outcomes-reach-a-ceiling}}

All 64 runs passed their hidden acceptance checks: every seeded bug was
actually fixed (not patched around), every mutation check was killed by
the new tests, every refactor preserved behavior, and every
documentation change matched the code. There were \textbf{zero scope
violations} in 64 runs: no agent in either condition touched a file
outside the task's expected set, no forbidden action occurred, and no
frozen test was modified. The baseline condition was not sloppier in
\emph{what it did}; both conditions were objectively correct (H3: no
room to improve).

Small tasks with capable models hit the anticipated triviality ceiling,
which sharpens the question the rest of the results answer: if the work
is equally correct, what does the contract buy?

\hypertarget{reviewability-what-the-contract-buys}{%
\subsection{Reviewability: what the contract
buys}\label{reviewability-what-the-contract-buys}}

Table 1 reports run-level paired results (median of three blinded
reviewers per run, 30 pairs). Rubric metrics are scored 1-5 unless
noted; decision is ordinal 0-3, accept is the clean-accept judge
fraction, and repair is binary.

\begin{longtable}[]{@{}
  >{\raggedright\arraybackslash}p{(\columnwidth - 12\tabcolsep) * \real{0.1429}}
  >{\raggedright\arraybackslash}p{(\columnwidth - 12\tabcolsep) * \real{0.1429}}
  >{\raggedright\arraybackslash}p{(\columnwidth - 12\tabcolsep) * \real{0.1429}}
  >{\raggedright\arraybackslash}p{(\columnwidth - 12\tabcolsep) * \real{0.1429}}
  >{\raggedright\arraybackslash}p{(\columnwidth - 12\tabcolsep) * \real{0.1429}}
  >{\raggedright\arraybackslash}p{(\columnwidth - 12\tabcolsep) * \real{0.1429}}
  >{\raggedright\arraybackslash}p{(\columnwidth - 12\tabcolsep) * \real{0.1429}}@{}}
\toprule\noalign{}
\begin{minipage}[b]{\linewidth}\raggedright
Metric
\end{minipage} & \begin{minipage}[b]{\linewidth}\raggedright
Baseline
\end{minipage} & \begin{minipage}[b]{\linewidth}\raggedright
Contract
\end{minipage} & \begin{minipage}[b]{\linewidth}\raggedright
Δ (B−A)
\end{minipage} & \begin{minipage}[b]{\linewidth}\raggedright
pairs +/−
\end{minipage} & \begin{minipage}[b]{\linewidth}\raggedright
Wilcoxon p
\end{minipage} & \begin{minipage}[b]{\linewidth}\raggedright
Cliff's \(\delta\)
\end{minipage} \\
\midrule\noalign{}
\endhead
\bottomrule\noalign{}
\endlastfoot
Evidence & 3.90 & 4.73 & \textbf{+0.83} & 22 / 0 & \textless{} 0.0001 &
0.66 \\
Ambiguity & 1.30 & 1.07 & \textbf{−0.23} & 2 / 9 & 0.035 & −0.23 \\
Confidence & 4.73 & 4.93 & +0.20 & 8 / 2 & 0.058 & 0.20 \\
Scope & 5.00 & 4.87 & −0.13 & 0 / 4 & 0.046† & −0.13 \\
Decision & 2.93 & 2.97 & +0.03 & 2 / 1 & 0.56 & 0.03 \\
Accept & 0.86 & 0.93 & +0.08 & 8 / 2 & 0.12 & 0.20 \\
Repair & 0.00 & 0.00 & 0.00 & 0 / 0 & n/a & 0.00 \\
\end{longtable}

† Only 4 non-zero pairs; the sign test gives p = 0.125. We treat this as
a hint rather than a finding.

\textbf{Evidence sufficiency (H2) improved in 22 of 30 pairs and never
worsened.} At the judge level, 62 of 90 paired reviewer scores favored
the contract and 0 favored the baseline (p \textless{} 0.0001). Reviewer
notes attribute the gap to exactly the elements the contract demands:
before/after test counts, per-file change reasons, explicit verification
steps, and stated limitations.

\textbf{Ambiguity (H1) fell from an already low baseline.} Most baseline
runs already scored the best possible ambiguity (1) because the tasks
were small; the contract removed most of the remaining uncertainty, with
9 pairs improved and 2 worsened (judge-level 22 vs.~3, p = 0.0001).

\textbf{The scope-compliance wrinkle.} The only metric that moved
\emph{against} the contract was scope compliance, by a small and
statistically fragile margin. Reviewer notes show the mechanism:
contracts elicited \emph{more} artifacts (a more thorough test file, a
new test module rather than an edit to an existing one), which blinded
reviewers occasionally read as mild over-delivery relative to the
minimal ask. Mechanically, none of these runs left the allowlist.
Explicit contracts shift output past some reviewers' minimal-change
expectations, which is a calibration question for contract authors
(state whether thoroughness beyond the minimum is wanted).

\hypertarget{work-package-structure}{%
\subsection{Work-package structure}\label{work-package-structure}}

In this pilot, the contract largely determines the shape of what comes
back:

\begin{longtable}[]{@{}llll@{}}
\toprule\noalign{}
Report element present & Baseline & Contract & Evidence bundle \\
\midrule\noalign{}
\endhead
\bottomrule\noalign{}
\endlastfoot
Changed files listed with reasons & 7\% & 93\% & 100\% \\
Commands run reported & 7\% & 70\% & 100\% \\
Known limitations stated & 0\% & 80\% & 100\% \\
Residual risks stated & 0\% & 0\% & 100\% \\
Reviewer checklist included & 0\% & 0\% & 100\% \\
Mean report length (chars) & 769 & 1,661 & 2,716 \\
\end{longtable}

The counts show two patterns. First, baseline agents succeeded at the
work just as often, but almost never volunteer the evidence a reviewer
needs; they report the fix, not the review surface. Second, some
evidence fields appear \emph{only on demand}: no agent in any of the 60
A/B runs spontaneously wrote a residual-risk section or a reviewer
checklist, while 100\% produced them when the evidence bundle required
it. Evidence is supplied elastically with respect to the contract, not
the task.

The four-run evidence-bundle preview scored at the top of the scale,
with evidence sufficiency 5.0 and ambiguity 1.0 on all four runs, versus
4.0/1.25 for matched baseline cells. That is enough to carry condition C
forward as a full arm, but not enough to estimate its effect.

\hypertarget{cost-of-the-contract-h4}{%
\subsection{Cost of the contract (H4)}\label{cost-of-the-contract-h4}}

Contracts added measurable run cost:

\begin{longtable}[]{@{}lllll@{}}
\toprule\noalign{}
Cost measure & Baseline & Contract & Overhead & Wilcoxon p \\
\midrule\noalign{}
\endhead
\bottomrule\noalign{}
\endlastfoot
Agent tokens per run & 20,087 & 22,690 & +13.0\% & 0.0002 \\
Wall-clock per run (s) & 32.7 & 45.2 & +38.3\% & \textless{} 0.0001 \\
Tool invocations & 7.9 & 9.7 & +23.3\% & 0.0012 \\
Patch size (bytes) & 1,556 & 2,252 & +44.7\% & n/a \\
\end{longtable}

The patch-size increase is dominated by additional tests, not production
churn (production files changed: 1.57 vs.~1.70 per run). On these tasks,
this overhead bought evidence and review surface, not correctness.
Whether that trade is worth it depends on the acceptance context: for
work that must be reviewed by someone other than its author (the
delegation setting by definition), evidence is the product.

\hypertarget{model-tier-and-task-family-effects}{%
\subsection{Model tier and task-family
effects}\label{model-tier-and-task-family-effects}}

The contract effect was roughly twice as large for the weaker tier.
Haiku pairs (n = 10): evidence +1.00, ambiguity −0.50, confidence +0.40.
Sonnet pairs (n = 20): +0.75, −0.10, +0.10. The stronger model
spontaneously writes better reports, so the contract has less left to
add. That pattern fits the interpretation that delegation contracts
partly substitute for unprompted reporting discipline, which matters
when delegating to cheaper or weaker executors.

By family, effects were largest for the most casual baseline asks,
namely docs updates and the simple clamp fix (evidence +1.33), and
smallest-to-negative for the duplication refactor (T07: ambiguity +0.67,
confidence −0.67), where the contract's ``the logic must exist exactly
once'' criterion brought interface-design judgment calls such as whether
the shared helper should also compute the pagination offset into focus
for reviewers.

\hypertarget{what-the-review-layer-catches}{%
\subsection{What the review layer
catches}\label{what-the-review-layer-catches}}

Three episodes illustrate what the review layer actually catches. (1)
One baseline report claimed ``all 29 tests now pass, including the
\textbf{four} auth tests that were previously failing'' when two had
been failing; two of three blinded reviewers flagged the inconsistency
and discounted the run's evidence score; report inaccuracy is a
reviewability defect even when the patch is perfect. (2) One contract
run disclosed that its first test version contained a faulty assertion
it found and fixed before submitting; reviewers explicitly credited the
disclosure. (3) One reviewer mis-attributed a detail across items in its
batch (claiming an import inconsistency that belongs to a different
task's baseline); the median over three independent reviewers absorbed
the error; single-reviewer scoring, human or model, would not have.

\hypertarget{discussion}{%
\section{Discussion}\label{discussion}}

\textbf{Contracts buy reviewability, not correctness, on tasks like
these.} The dissociation is the central result. With capable models and
small, well-specified tasks, the marginal return of an explicit contract
on \emph{outcome} is zero because outcomes saturate; the return on
\emph{evidence} is large, uniform (no pair got worse), and concentrated
exactly in the fields the contract names. The conceptual claim that the
delegation contract is a control and review layer {[}4{]} survives its
first measurement in a specific, falsifiable form: the contract controls
the work package more than the work.

\textbf{Evidence is demand-elastic.} The clearest structural result is
0\% → 100\%: residual-risk sections and reviewer checklists never appear
spontaneously and always appear on demand. Agents comply with evidence
requirements nearly perfectly; they just do not anticipate them. The
delegating system therefore has to ask for review evidence explicitly.

\textbf{Implications for harness builders.} If a harness's job includes
making delegated work reviewable, these defaults follow directly: (i)
require a changed-files-with-reasons list and a tests-run section in
every agent report (cheap, near-universal compliance, largest evidence
gain); (ii) require known-limitations and residual-risk sections for
anything above trivial risk, since they will not appear otherwise; (iii)
expect \textasciitilde15\% token and \textasciitilde40\% latency
overhead and surface it as the price of evidence; (iv) when delegating
to weaker/cheaper models, contracts matter more, not less; (v) state in
the contract whether thoroughness beyond the minimal ask is welcome, to
avoid penalizing over-delivery at review time.

\textbf{Negative space matters.} The pilot also says what contracts did
\emph{not} do here: they did not reduce scope violations (there were
none to reduce), did not change accept/reject decisions on correct work,
and never triggered the repair-required flag. A study that wants to
measure those margins needs tasks hard enough for baseline runs to fail,
drift, or game tests; that is the next iteration's defining requirement.

\hypertarget{threats-to-validity}{%
\section{Threats to Validity}\label{threats-to-validity}}

\textbf{Model-based reviewers.} All 192 reviews were produced by LLM
reviewers, not humans. LLM-as-judge protocols correlate imperfectly with
human judgment {[}8{]}, and reviewer errors occurred (Section 4.6),
though median-of-three aggregation contained them. The mechanical layer
(hidden tests, mutation checks, scope analysis) is model-free and
unaffected. Human-reviewer replication on a subset is the most important
follow-up.

\textbf{In-family bias.} Agents and reviewers are different deployments
of the same model family (and the study itself was orchestrated by a
related model; see Disclosure). Reviewers may share stylistic
preferences with the agents whose reports they score. The paired design
removes any constant bias between conditions, but family-specific taste
could inflate absolute scores.

\textbf{Partial blinding.} Condition labels were hidden and order
randomized, but a templated report is recognizably templated, so full
blinding of structure is impossible when structure is the treatment. The
mechanical metrics and the within-rubric anchors mitigate, but do not
eliminate, this.

\textbf{Task triviality.} Objective metrics hit the design-anticipated
ceiling. Findings generalize to ``small, well-specified tasks,'' not to
risky migrations or large features; H3 (authority control) remains
untested.

\textbf{Single repository, single language, two related models.} The
study uses one \textasciitilde600-line TypeScript API; results may
differ for larger codebases, other languages, other model families, and
other harnesses.

\textbf{No review-time measurement.} H4's ``total acceptance cost'' is
only partially estimated: run cost was measured, while review effort was
proxied by rubric scores rather than minutes.

\textbf{Small C arm.} The evidence-bundle condition ran as a four-run
preview, and its perfect scores are encouraging but not load-bearing.

\hypertarget{related-work}{%
\section{Related Work}\label{related-work}}

Repository-level agent evaluation (SWE-bench {[}2{]}, SWE-agent {[}3{]},
Agentless {[}9{]}) measures whether agents can resolve issues; this
study holds resolution fixed and measures the review layer around it.
Mined corpora of agent-authored pull requests (AIDev {[}5{]}) document
delegated work at scale but cannot reconstruct the original prompts,
authority boundaries, or evidence requirements, which are precisely the
variables manipulated here. Supervisory-control theory {[}6{]} and
meaningful-human-control accounts {[}10{]} supply the framing that
delegation is a control relationship; human-AI teaming results {[}7{]}
argue that team value depends on verification cost, which our
evidence-sufficiency results operationalize. Studies of provenance and
disclosure {[}11, 14{]} show that knowing how code was produced changes
reviewer behavior; we complement them by showing that \emph{contracting
for evidence} changes what there is to review. Software-bot surveys
{[}15{]} and automated program repair {[}16{]} anticipate the
artifact-with-evidence framing. The delegation-contract framework
itself, including the \(\langle T, A, W, C\rangle\) model and the
testing agenda this pilot executes, is developed in {[}4{]}.

\hypertarget{conclusion}{%
\section{Conclusion}\label{conclusion}}

We turned the software-delegation-contract framework into a measurement
instrument and ran it. In 64 controlled coding-agent runs, making the
delegation contract explicit did not change whether the work succeeded.
Success saturated, but reviewability changed reliably: evidence
sufficiency improved, reviewer ambiguity decreased, and evidence fields
appeared on demand even when they never appeared spontaneously. The cost
was roughly 13\% more tokens and 38\% more wall-clock time. A software
delegate is only as useful as its work is reviewable; the delegation
contract is one practical way to make that review possible.

A larger study should change three parts of the design: use tasks hard
enough for baseline runs to fail or drift, add human reviewers on at
least a subset, and run the evidence-bundle condition as a full arm. The
complete harness is preserved as a companion evidence artifact,
including the seeded task repository, paired prompts, hidden acceptance
and mutation checks, blinded review pipeline, and analysis scripts.

\hypertarget{disclosure}{%
\section{Disclosure}\label{disclosure}}

The author develops AI-assisted software tooling, but no specific
product is described or evaluated. The study was executed end-to-end by
an AI research assistant (a Claude-family model operating the Claude
Code harness) under the author's prior written study design: it
generated the task environment, ran the agent conditions, performed the
blinded model-based reviews, computed the statistics, and drafted this
manuscript. All prompts, patches, work packages, reviews, and scripts
are preserved verbatim for audit and later independent verification. The
coding agents under test, the reviewers, and the orchestrating assistant
are different deployments of the same model family; the implications are
discussed in Threats to Validity.

\hypertarget{references}{%
\section{References}\label{references}}

\begin{enumerate}
\def\labelenumi{\arabic{enumi}.}
\tightlist
\item
  OpenAI Codex Documentation. ``Codex web''.
  \url{https://developers.openai.com/codex/cloud}. Accessed June 12,
  2026.
\item
  Carlos E. Jimenez, John Yang, Alexander Wettig, Shunyu Yao, Kexin Pei,
  Ofir Press, and Karthik Narasimhan. ``SWE-bench: Can Language Models
  Resolve Real-World GitHub Issues?'' arXiv:2310.06770, 2023.
\item
  John Yang, Carlos E. Jimenez, Alexander Wettig, Kilian Lieret, Shunyu
  Yao, Karthik Narasimhan, and Ofir Press. ``SWE-agent: Agent-Computer
  Interfaces Enable Automated Software Engineering''. arXiv:2405.15793,
  2024.
\item
  Vincent Schmalbach. ``Software Delegates: Delegation Contracts for AI
  Coding Agents''. Working paper, 2026.
  https://www.vincentschmalbach.com/.
\item
  Hao Li, Haoxiang Zhang, and Ahmed E. Hassan. ``AIDev: Studying AI
  Coding Agents on GitHub''. arXiv:2602.09185, 2026.
\item
  Raja Parasuraman, Thomas B. Sheridan, and Christopher D. Wickens. ``A
  model for types and levels of human interaction with automation''.
  IEEE Transactions on Systems, Man, and Cybernetics, Part A,
  30(3):286--297, 2000.
\item
  Gagan Bansal, Besmira Nushi, Ece Kamar, Eric Horvitz, Daniel S. Weld,
  and Walter S. Lasecki. ``Updates in Human-AI Teams: Understanding and
  Addressing the Performance/Compatibility Tradeoff''. AAAI, 2019.
\item
  Lianmin Zheng, Wei-Lin Chiang, Ying Sheng, et al.~``Judging
  LLM-as-a-Judge with MT-Bench and Chatbot Arena''. arXiv:2306.05685,
  2023.
\item
  Chunqiu Steven Xia, Yinlin Deng, Soren Dunn, and Lingming Zhang.
  ``Agentless: Demystifying LLM-based Software Engineering Agents''.
  arXiv:2407.01489, 2024.
\item
  Filippo Santoni de Sio and Jeroen van den Hoven. ``Meaningful Human
  Control over Autonomous Systems: A Philosophical Account''. Frontiers
  in Robotics and AI, 2018.
\item
  Ningzhi Tang, Meng Chen, Zheng Ning, Aakash Bansal, Yu Huang, Collin
  McMillan, and Toby Jia-Jun Li. ``A Study on Developer Behaviors for
  Validating and Repairing LLM-Generated Code Using Eye Tracking and IDE
  Actions''. arXiv:2405.16081, 2024.
\item
  GitHub Docs. ``About GitHub Copilot cloud agent''.
  \url{https://docs.github.com/en/copilot/concepts/agents/cloud-agent/about-cloud-agent}.
  Accessed June 12, 2026.
\item
  Claude Code Documentation. ``Claude Code overview''.
  \url{https://code.claude.com/docs/en/overview}. Accessed June 12,
  2026.
\item
  Syed Mohammad Kashif, Peng Liang, and Amjed Tahir. ``On Developers'
  Self-Declaration of AI-Generated Code: An Analysis of Practices''.
  arXiv:2504.16485, 2025.
\item
  Sivasurya Santhanam, Tobias Hecking, Andreas Schreiber, and Stefan
  Wagner. ``Bots in software engineering: a systematic mapping study''.
  PeerJ Computer Science, 8:e866, 2022.
\item
  Martin Monperrus. ``Automatic Software Repair: A Bibliography''.
  arXiv:1807.00515, 2018.
\end{enumerate}

\end{document}